\documentclass[a4paper,fleqn,usenatbib]{mnras}

\usepackage[T1]{fontenc}
\usepackage{ae,aecompl}

\usepackage[normalem]{ulem}
\usepackage{amsmath}
\usepackage{graphicx} 
\usepackage{lscape}
\usepackage{indentfirst}
\usepackage{enumitem}
\usepackage{xspace}

\usepackage{amssymb}
\usepackage[dvipsnames]{xcolor}
\usepackage{url}
\usepackage[flushleft]{threeparttable}

\newcommand {\bc}{\begin {center}}
\newcommand {\ec}{\end {center}}
\newcommand {\be}{\begin {equation}}
\newcommand {\ee}{\end {equation}}
\newcommand {\beq}{\begin {eqnarray}}
\newcommand {\eeq}{\end {eqnarray}}
\newcommand {\comment}[1]{}

\renewcommand{\d}{{\rm d}}

\newcommand {\ergs}{{\rm erg\ \rm s^{-1}}}

\title[Radiation/accretion flow coupling in ULX pulsars]
{
Coupling of radiation and magnetospheric accretion flow in ULX pulsars: radiation pressure and photon escape time
}
\author[C.~Flexer and A.~A.~Mushtukov] 
{
Caitlyn~Flexer$^{1}$\thanks{E-mail: caitlyn.flexer@gmail.com (CF)}
and
Alexander~A.~Mushtukov$^{2}$\thanks{E-mail: alexander.mushtukov@physics.ox.ac.uk (AAM)}
\\ 
$^1$ Castilleja School, 1310 Bryant St, Palo Alto, CA 94301 Palo Alto, USA  \\
$^2$ Astrophysics, Department of Physics, University of Oxford, Denys Wilkinson Building, Keble Road, Oxford OX1 3RH, UK \\ 
} 

\pubyear{2023}

\begin{document}
\label{firstpage}
\pagerange{\pageref{firstpage}--\pageref{lastpage}}
\maketitle


\begin{abstract} 
The accretion flow within the magnetospheric radius of bright X-ray pulsars can form an optically thick envelope, concealing the central neutron star from the distant observer. 
Most photons are emitted at the surface of a neutron star and leave the system after multiple reflections by the accretion material covering the magnetosphere. 
Reflections cause momentum to be transferred between photons and the accretion flow, which contributes to the radiative force and should thus influence the dynamics of accretion. 
We employ Monte Carlo simulations and estimate the acceleration along magnetic field lines due to the radiative force as well as the radiation pressure across magnetic field lines.
We demonstrate that the radiative acceleration can exceed gravitational acceleration along the field lines, and similarly, radiation pressure can exceed magnetic field pressure.
Multiple reflections of X-ray photons back into the envelope tend to amplify both radiative force along the field lines and radiative pressure.
We analyze the average photon escape time from the magnetosphere of a star and show that its absolute value is weakly dependent on the magnetic field strength of a star and roughly linearly dependent on the mass accretion rate being $\sim 0.1\,{\rm s}$ at $\dot{M}\sim 10^{20}\,{\rm g\,s^{-1}}$. 
At high mass accretion rates, the escape time can be longer than free-fall time from the inner disc radius.
\end{abstract}


\section{Introduction}
\label{sec:Intro}

X-ray pulsars (XRPs) are strongly magnetized, accreting neutron stars (NSs) in close binary systems (see \citealt{2022arXiv220414185M} for review).
The typical magnetic field strength at the NS surface in XRPs is estimated to be $\sim 10^{12}\,{\rm G}$, which was determined from the detection of cyclotron lines in the X-ray energy spectra of some objects (see \citealt{2019A&A...622A..61S} for review).
The apparent luminosity of XRPs covers about nine orders of magnitude from $\sim 10^{32}\,\ergs$ to $\sim 10^{41}\,\ergs$. 
The lower limit of accretion luminosity is due to the onset of the centrifugal barrier and the ``propeller" effect of accretion \citep{1975A&A....39..185I,2016A&A...593A..16T}.
The brightest XRPs have luminosities that exceed the Eddington limit for a NS:
$
L_{\rm Edd}\simeq {4\pi GM m_{\rm p}c}/[\sigma_{\rm T} (1+{\rm x}_H)]\simeq 2\times 10^{38}(M/M_\odot)\,\ergs
$, 
where $m_{\rm p}$ is a mass of proton, $\sigma_{\rm T}$ is the Thomson scattering cross-section, and ${\rm x}_H$ is the hydrogen mass fraction in accreting material. These XRPs belong to the recently discovered class of pulsating ultra-luminous X-ray sources (ULXs, see e.g.,  \citealt{2021AstBu..76....6F,2023NewAR..9601672K}).
Currently, six pulsating ULXs have been detected: 
M82~X-2 \citep{2014Natur.514..202B},
NGC~7793~P13 \citep{2016ApJ...831L..14F,2017MNRAS.466L..48I},
NGC~5907~X-1 \citep{2017Sci...355..817I},
NGC~300~X-1 \citep{2018MNRAS.476L..45C},
M51~X-7 \citep{2020ApJ...895...60R},
NGC~1313~X-2 \citep{2019MNRAS.488L..35S}.
Luminosity values comparable to the luminosity of ULX pulsars can be detected in some Be X-ray transients at the peaks of the outbursts (see \citealt{2011Ap&SS.332....1R} for review). 

The strong magnetic field shapes the geometry of accretion flow in XRPs.
The accretion flow in form of stellar wind or accretion disc is interrupted by the strong magnetic field at a distance called the magnetospheric radius, where the accretion material starts to move along magnetic field lines towards NS surface.
By comparing the ram pressure of accreting material $P_{\rm ram}\propto \rho v^2$ (where $\rho$ and $v$ are material mass density and velocity) and the pressure of the NS magnetic field $P_B\propto B^2\propto B^2_0\,r^{-6}$ (where $B$ is magnetic field strength at distance $r$ from a NS and $B_0$ is field strength at the NS surface at the NS surface), we can estimate the magnetospheric radius (see, e.g., Chapter 6.3 in \citealt{2002apa..book.....F}).
In the case of a magnetic field dominated by the dipole component, the magnetospheric radius can be estimated as
\beq \label{eq:Rm}
R_{\rm m} \simeq 5\times 10^7\,\Lambda B_{12}^{4/7}\dot{M}_{19}^{-2/7}m^{-1/7}R_6^{12/7}\,\,{\rm cm},   
\eeq 
where $\Lambda\leq 1$ is dimensionless coefficient, $B_{12}$ is the field strength at the NS magnetic pole in units of $10^{12}\,{\rm G}$, $\dot{M}_{19}$ is the mass accretion rate in units of $10^{19}\,{\rm g\,s^{-1}}$, $m$ is the mass of a NS in units of Solar masses, and $R$ is NS radius in units of $10^6\,{\rm cm}$.
The coefficient $\Lambda$ is typically taken to be $\Lambda=0.5$ for the case of accretion from the disc and $\Lambda=1$ for the case of accretion from the wind \citep{1978ApJ...223L..83G,1979ApJ...234..296G}.
However, in the case of accretion from the disc, it is dependent on the mass accretion rate and physical conditions at the inner disc radius
(see \citealt{1999ApJ...521..332P,2019A&A...626A..18C} for details and discussion).
At the magnetospheric radius, the accretion flow penetrates into the NS magnetosphere due to the development of instabilities, including reconnection, magnetic Rayleigh-Taylor \citep{1976ApJ...207..914A,2008MNRAS.386..673K} and Kelvin-Helmholtz  instabilities \citep{1983ApJ...266..175B}.
Within the magnetosphere of a NS, material moves toward the NS along magnetic field lines and finally reaches the stellar surface in a small regions (area $\sim 10^9\,{\rm cm^2}$) near the magnetic poles of a NS.
The kinetic energy of the accretion flow turns into heat and is emitted in the form of X-rays by hot spots or accretion columns, depending on the mass accretion rate and surface magnetic field strength and structure \citep{1976MNRAS.175..395B,2015MNRAS.447.1847M}.

The NS magnetic field in XRPs is typically assumed to be dominated by the dipole component.
However, evidence of strong non-dipole magnetic field was reported in a few objects including Her~X-1 \citep{1991PAZh...17..803S,2022MNRAS.515..571M}, bright Be X-ray transients SMC~X-3 \citep{2017A&A...605A..39T}
and Swift~J0243.6+6124 \citep{2022ApJ...933L...3K},
and ULX pulsar NGC~5907~X-1 \citep{2017Sci...355..817I}.

Typically, it is assumed that the accretion flow between the inner disc radius and NS surface does not modify the X-ray flux and spectra in XRPs because the magnetospheric flow is optically thin.
\citealt{1976SvAL....2..111S}, however, noticed that even at low mass accretion rates, the flow within the magnetosphere of a NS can absorb and reprocess photons of low energies, which can result in relatively low pulsed fraction observed in soft X-ray energy band.
At high mass accretion rates $\dot{M}\gtrsim 10^{18}\,{\rm g\,s^{-1}}$, the accretion flow inside the magnetospheric radius is expected to be optically thick \citep{2017MNRAS.467.1202M} due to the Compton scattering.
Under this condition, accretion should affect the processes of spectra formation, pulse profiles and features of the aperiodic variability \citep{2019MNRAS.484..687M,2019A&A...621A.118K,2023MNRAS.525.4176B}.
In particular, optically thick envelopes in ULX pulsars can shield central NS from an observer, reprocess a large fraction of X-ray flux and re-emit it in softer multi-color blackbody radiation \citep{2019A&A...621A.118K,2023MNRAS.525.4176B}.
\citealt{2023MNRAS.525.4176B} demonstrated that appearance of optically thick envelopes in XRPs results in one-peak pulse profile instead of pulse profile with two peaks of lower mass accretion rates. 
The numerous absorptions and re-emissions of X-ray photons by the envelope contributes to the typical escape time of photons from the system.
Therefore, one would expect suppression of X-ray aperiodic variability on time scales smaller than the typical escape time \citep{2019MNRAS.484..687M}. 

The dynamics of the flow covering NS magnetosphere  is governed by the geometry of magnetic field lines, gravitational force, centrifugal force and, in the case of high mass accretion rates and luminosities, the radiative force.
The flow dynamics within the magnetospheric radius was already simulated numerically by \citealt{2003ApJ...595.1009R,2004ApJ...610..920R}, but these simulations were limited to the case of low mass accretion rates (where one can safely neglect the influence of radiative force) and relatively small magnetospheres of a NS.
Self-consistent simulations of accretion flow dynamics in bright XRPs requires calculations of radiative force, which was not performed up to date.

In this paper, we investigate the radiative force applied to the material covering the magnetosphere of a NS, the role of multiple reflections of X-ray photons, and the typical time scale of photon escape from the magnetosphere covered by an optically thick accretion flow.
To estimate the radiative force, we use pre-calculated maps of material distribution over the magnetospheric surface, and perform Monte Carlo simulations under the assumption that the magnetic field geometry is dominated by the dipole component, while the opacity is due to the scattering only.


\section{Model set up}
\label{sec:Model}

\begin{figure}
\centering 
\includegraphics[width=8.2cm]{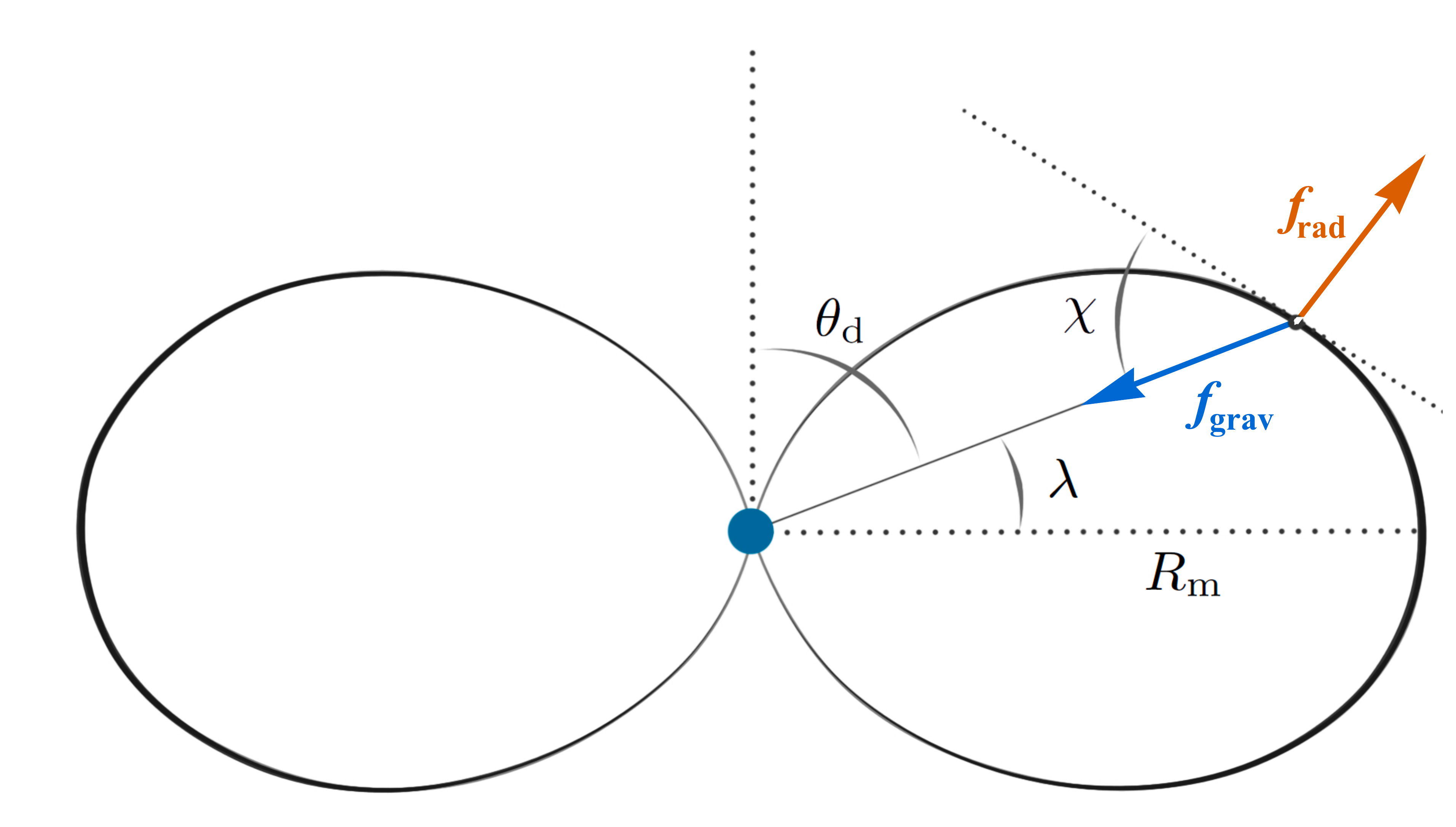}
\caption{
The schematic illustration of dipole magnetic field lines geometry.
In general, the gravitational and the radiative forces applied to a given point at the magnetospheric surface are not strictly in opposite directions due to the possibility of multiple photon reprocessing by the accretion flow.
}
\label{pic:scheme}
\end{figure}

We consider a NS with mass $M=1.4 M_\odot$ and radius $R=10^6\,{\rm cm}$ in a binary system. 
We assume that the accretion flow forms a disc outside of the magnetospheric radius (\ref{eq:Rm}).
The magnetic and rotational axis of a NS are assumed to be orthogonal to the disc plane.
At a distance $R_{\rm m}$ from the central object, the accretion flow is interrupted by the magnetic field of the NS.
Starting from the inner disc radius, the accretion flow follows magnetic field lines and moves towards NS surface, forming an envelope that covers the magnetosphere of the star.
In the case of strongly magnetised NSs, $R_{\rm m}\gg R$ and the majority of energy is released at the NS surface.
The luminosity related to the mass accretion rate is 
\beq 
L=\frac{GM\dot{M}}{R}\simeq 1.3\times 10^{39}\dot{M}_{19}m R_6^{-1}\,\ergs.
\eeq 
Due to the cylindrical symmetry, the coverage of the magnetosphere with matter is expected to be independent of the azimuthal angle. 
The motion of material along magnetic field lines starts in the equatorial plane with the velocity close to the thermal velocity in the accretion disc.
Estimating the inner disc radius, we assume that the parameter $\Lambda$ in (\ref{eq:Rm}) is not dependent on the mass accretion rate and equals $0.5$. 
However, we account for the dependence of the accretion disc geometrical thickness at the inner radius on the mass accretion rate.
Estimating the geometrical thickness of a disc we follow \citealt{2007ARep...51..549S} and get
\beq\label{SS_H_c}
H_{\rm d}\approx 2.7\times 10^6 \,\alpha^{-1/10}\Lambda^{9/8}\, \dot{M}_{19}^{-6/35} B_{12}^{9/14}\, m^{-15/28}\,R_6^{27/14}\,{\rm cm}
\eeq
for the case of relatively low mass accretion rates
$$\dot{M}_{19}\lesssim 0.04\,\Lambda^{21/20}m^{-1/2}R_6^{9/5},$$ 
when the disc is gas pressure dominated and the opacity is given by the Kramers’ law.
At higher mass accretion rates, the disc is assumed to be radiation pressure dominated with the opacity determined by the electron scattering.
In this case, the geometrical thickness of the disc is expected to be independent of the radial coordinate \citep{1973A&A....24..337S} and can be estimated as
\beq
H_{\rm d}\approx 1.3\times 10^{7}\,\dot{M}_{19}\,\mbox{cm}.
\eeq

We consider the magnetic field of a NS dominated by a dipole component (see Fig.\,\ref{pic:scheme}). 
The geometry of magnetic field lines in spherical coordinates is described by
\beq 
r (\theta,\varphi) = R_{\rm m}\sin^2\theta = 
R_{\rm m}\cos^2\lambda .
\eeq 
The surface area of a segment on the dipole surface is given by
\beq
\d S=R_{\rm m}^2 \cos^4\lambda (1+3\sin^2\lambda)^{1/2} \d\lambda\d\varphi,
\eeq
while the length along the field lines $x$ is related to the latitude:
\beq
\d x = R_{\rm m}\cos\lambda (1+3\sin^2\lambda)^{1/2} \d\lambda.
\eeq
The angle between the position vector of the point at the dipole surface and the tangent line to the dipole magnetic field line (see Fig.\,\ref{pic:scheme}) is
\beq
\chi={\rm atan}\left(\frac{0.5}{{\rm tan}\lambda}\right).
\eeq
The magnetic field strength in the case dominant dipole component drops with the distance as $B\propto r^{-3}$ and the field strength at the magnetosheric radius can be estimated as 
\beq 
B(r=R_{\rm m}) \approx 3\times 10^6\,\Lambda^{-3}B_{12}^{-5/7}L_{39}^{6/7}m^{-3/7}R_6^{-9/7}\,\,{\rm G}.
\eeq 
Note that the stronger the field strength at the NS surface, the smaller it is at the magnetospheric radius.

The accretion flow moves across the magnetospheric surface strictly along magnetic field lines under the influence of gravitational, centrifugal and the radiative forces. 
The magnitude of acceleration due to the gravitational force along the magnetic field lines is
\beq
\left|a_{{\rm grav},||}\right|=\cos\chi\,\frac{GM}{r^2}\simeq 1.33\times 10^{10}\cos\chi\,\frac{m}{r_8^2} \,\,{\rm cm\,s^{-2}}.
\eeq
In the case of dipole magnetic field axis orthogonal to the disc plane, the acceleration due to the centrifugal force is  
\beq\label{eq:acc_centr_1}
a_{{\rm cen},||}(\alpha=0)=\omega^2 R_{\rm m}\cos^3\lambda \cos(\chi-\lambda),
\eeq
where $\omega=2\pi/P_{\rm spin}$ is the angular velocity of a NS, and $\alpha$ is the angle between the magnetic and rotation axis of a NS.

The radiative force is dependent on the luminosity of the central object, the beam pattern, the surface density of the material covering the magnetosphere, the optical thickness, and details of the radiative transfer (i.e., the dominant processes of radiation and matter interaction).
We assume that the opacity of matter covering the magnetospheric surface is due to the scattering of photons by electrons (Compton scattering), and the opacity is 
\beq
\kappa_e=0.34\,{\rm cm^2\,g^{-1}}.
\eeq 
Free-free absorption can be approximatelly described by the Kramers formula:
\beq \kappa_{\rm ff}\approx
0.0136\,\rho T_{\rm keV}^{-3.5}\,\,\,{\rm cm^2 \,g^{-1}},
\eeq
because the temperature of the magnetospheric accretion flow is expected to be $T\sim 1\,{\rm keV}$ and the mass density in accretion channel $\rho\ll 1\,{\rm g\,cm^{-3}}$, $\kappa_{\rm ff}\ll\kappa_e$.

If one considers only scattering and neglects the absorption of photons in the magnetospheric flow, each photon that reaches the magnetosphere either 
\begin{enumerate}[leftmargin=*]
\item 
passes through the layer without interacting with it, 
\item
passes through the layer after several scatterings, or 
\item
is reflected back inside the magnetosphere, having scattered once or multiple times.
\end{enumerate}
The fraction of photons that pass through the layer without interacting with the accretion flow is 
$$f = e^{-\tau_e / \cos\theta_{\rm in}},$$
where $\theta_{\rm in}$ is the angle between the momentum of incident photon and local normal to the magnetospheric surface.
These photons do not transfer any momentum to the flow and thus do not contribute to the radiative force.
The photons passing through the layer after one or more scatterings contribute to the radiative force because the direction of their momentum changes after each reflection.
Moreover, the photons reflected back into the NS magnetosphere contribute to the radiative force after each reflection.
In the case of high mass accretion rates and optically thick accretion columns, photons can experience many reflections on average, and the reflection process can heavily influence the radiative force and radiative pressure.

\begin{figure}
\centering 
\includegraphics[width=8.5cm]{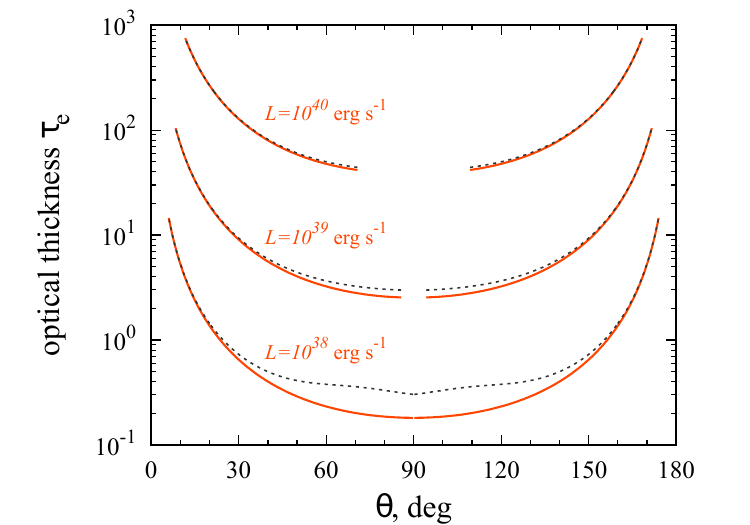}
\caption{
Optical thickness of accretion flow covering magnetosphere of a NS as a function of $\theta$-coordinate.
Different curves are given for different accretion luminosity: $10^{38}$ (lower), $10^{39}$ (middle), $10^{40}\,\ergs$ (upper).
Red solid curves are calculated for the case of non-rotating magnetosphere, while the black dashed ones account for NS rotation with spin period $P=1\,{\rm s}$.
Magnetic field at the NS surface is fixed at $B=3\times 10^{12}\,{\rm G}$.
Calculation are performed according \citealt{2017MNRAS.467.1202M,2019MNRAS.484..687M}.
}
\label{pic:tau}
\end{figure}

\begin{figure}
\centering 
    \includegraphics[width=8.5cm]{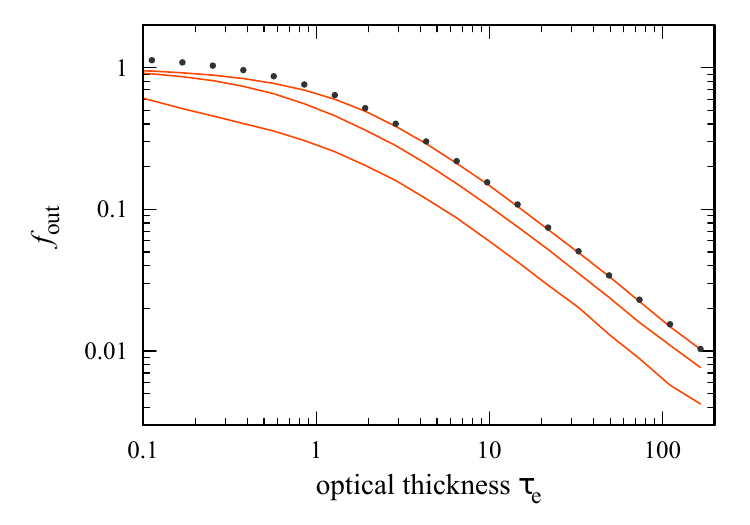}
\caption{
The fraction of photons that penetrate through the layer of optical thickness $\tau_e$.
Red lines show the fractions calculated for different initial angles between the photon momentum and normal to the layer: $\theta_{\rm in}=0\degr$, $60\degr$ and $86\degr$ (top to bottom, see Appendix\,\ref{app:fraction}).
Black dots show approximation calculated according to (\ref{eq:f_out_approx}).
}
\label{pic:sc_f_out}
\end{figure}

In the case of azimuthal symmetry of magnetospheric coverage by the accreting material, the local optical thickness of the accretion flow due to the scattering can be roughly estimated as
\beq\label{eq:tau_e}
\tau_e \approx \frac{70 L_{39}^{6/7}B_{12}^{2/7}}{\beta}\left(\frac{\cos\lambda_0}{\cos\lambda}\right)^3,
\eeq 
where $\beta=v/c$ is local dimensionless velocity (see Section\,2.1 in \citealt{2017MNRAS.467.1202M} and Fig.\,\ref{pic:tau} illustrating more accurate calculations), and $\lambda_0\approx{\pi}/{2}-(R/R_{\rm m})^{{1}/{2}}$ is the angle at the accretion channel base at the NS surface.
If the magnetospheric flow is optically thick, the photons that originated from the central source experience several reflections back into the NS magnetosphere before they finally leave the system.
The fraction of photons that penetrate through the layer is dependent on the optical thickness $\tau_e$ of a layer and initial angle between photon momentum and local normal (see Fig.\,\ref{pic:sc_f_out}).
For the case of an optically thick layer (i.e., $\tau_e\gtrsim 1$), where the radiative transfer is dominated by the isotropic scattering, and photons incident perpendicular to the surface (i.e., $\theta_{\rm in}=0$), the fraction of particles penetrating through the layer can be approximated by 
(see Appendix~C in \citealt{2019MNRAS.484..687M} and \citealt{2022Ap.....65..560N} for detailed analytical discussion)
\footnote{Note, that there were a typo in this approximate expression in \citealt{2019MNRAS.484..687M}.}
\beq\label{eq:f_out_approx}
f_{\rm out} \simeq \frac{\sqrt{3}}{\tau_e+1.42}.
\eeq 
For larger $\theta_{\rm in}$, the fraction of photons penetrating through the layer is smaller. 

The construction of a self-consistent model of magnetospheric accretion falls outside the scope of this paper. 
To assess the impact of multiple reprocessings of X-ray photons in the magnetosphere of a NS, we utilize precalculated maps of matter distribution over the magnetospheric surface. 
These maps are based on the models proposed by \citealt{2017MNRAS.467.1202M,2019MNRAS.484..687M} (examples of calculations are illustrated in Fig.,\ref{pic:tau}).

\section{Numerical model}
\label{sec:NumModel}

To calculate the radiative force and momentum exchange between the accretion flow and photons, we perform Monte Carlo simulations of momentum transfer to the accretion envelope.
Each simulation is composed of a few steps:
\begin{enumerate}[leftmargin=*]
\item
We start a photon from the central source. 
The initial direction of a photon is described by two angles - $\theta_i$ and $\varphi_i$.
{
The angles are determined by random numbers $X_1,X_2\in[0;1]$.
In the case of central source producing isotropic X-ray emission, the initial direction of photon momentum is given by 
\beq 
\theta_i^{\rm (iso)} &=& {\rm acos}(1-2X_1)\in \left[0;\pi\right],\\
\varphi_i^{\rm (iso)} &=& 2\pi X_2 \in [0;2\pi].\eeq 
Each photon in this case is assigned the same statistical weight $w_i=1$.
By changing the statistical weight $w_i$ of a photon depending on the initial direction of its motion we can take into account the possible non-isotropy of the radiation from the central NS.
In the particular case of 
\beq 
\frac{\d L}{\d\Omega_i}=  g(\theta_i,\varphi_i)L,
\eeq 
where $\Omega_i$ is the solid angle, $g$ is a function describing initial angular distribution of X-ray photons taken so that $\int_0^\pi\d\theta_i\int_0^{2\pi}\d\varphi_i g(\theta_i,\varphi_i) \sin\theta_i =1$,
the photons' statistical weight is taken as 
\beq\label{eq:ph_weight}
w_i=g(\theta_i,\varphi_i).
\eeq 
Each photon participating in the simulation carries effective momentum $p_i\propto w_i$, or more precisely 
\beq 
p_i = \frac{w_i}{N_{\rm tot}}\frac{L}{c},
\eeq 
where $N_{\rm tot}$ is the total number of photons used in a simulation.
}\\
\item\label{step:get_new_cross}
The photon started from a given coordinate in a certain direction is traced as it travels to the magnetospheric surface, under the assumption that it is propagating along a straight-line trajectory. 
We determine the coordinates of the point at the magnetospheric surface at which the photon eventually reaches.
\item 
Using (\ref{eq:tau_e}), we estimate the local optical thickness $\tau_{\rm e}$ of the accretion flow at the point where the photon reaches the magnetosphere. 
We use the local optical thickness and angle between the normal to the magnetospheric surface and photon momentum to calculate the probability $f_{\rm out}$ for the photon to penetrate through the envelope (see approximation \ref{eq:f_out_approx} and Fig.\,\ref{pic:sc_f_out}).
We generate a random number $X_3$ between 0 and 1 and compare it to $f_{\rm out}$ in order to determine whether the photon will penetrate through the envelope or reflect back into the dipole magnetosphere.
If the photon penetrates through the flow, we stop tracing its trajectory and proceed to step \ref{step:time_accounting}. 
If the photon is reflected back into the NS magnetosphere, we determine the new direction of its motion, account for the local momentum transfer to the material along and across magnetic field lines, and continue tracing its trajectory.
Each reflection is associated with momentum transfer between the accretion flow and X-ray photons. 
The momentum effectively transferred to the accretion flow due to a reflection is
\beq
\Delta\mathbfit{p} = \mathbfit{p}_{\rm ini} - \mathbfit{p}_{\rm fin},
\eeq
where $\mathbfit{p}_{\rm ini}$ and $\mathbfit{p}_{\rm fin}$ are vectors of the photon's momentum before and after the reflection respectively, and $w_i$ is a photon effective weight (\ref{eq:ph_weight}).
The new direction of photon motion after reflection is calculated under the assumption of multiple conservative, isotropic and coherent scattering in the layer of a given optical thickness.
We return to step \ref{step:get_new_cross}.
\item\label{step:time_accounting} 
We account for the photon time travel before its escape in our estimations of the average photon escape time.
\end{enumerate}

The momentum transferred from photons to the accretion flow determines the radiative force applied to the flow and the corresponding acceleration along magnetic field lines:
\beq\label{eq:a_par} 
a_{\rm rad, ||} \geq
a^*_{\rm rad, ||} = \left(\frac{\sum_{j}\Delta p_{j,||}}{N_{\rm tot}}\right)\frac{L}{c}\frac{1}{\Sigma \Delta S},
\eeq 
where $\Sigma$ is the local surface density of the accretion flow, and $\Delta S$ is the surface area of the segment.
The projection of the transferred momentum onto the direction of the local magnetic field lines is given by the scalar production
\beq\label{eq:momentum_par}
\Delta p_{j,||}=\Delta \mathbfit{p}_{j}\cdot \mathbfit{n}_b
\eeq
where 
\beq
\mathbfit{n}_b = 
\left(\begin{array}{c} 
-\cos(\chi-\lambda)\cos\varphi \\ 
-\cos(\chi-\lambda)\sin\varphi \\
\sin(\chi-\lambda)
\end {array}\right) 
\eeq 
is the unit vector tangential to the magnetic field lines. \\
Note that the estimation (\ref{eq:a_par}) assumes that the momentum transfer is distributed homogeneously over the geometrical thickness of the accretion flow, which is not necessarily the case.

The radiative pressure on the magnetosphere is related to the photon momentum transfer across magnetic field lines: 
\beq 
P_{\rm rad, \perp} = \left(\frac{\sum_{j}\Delta p_{j,\perp}}{N_{\rm tot}}\right)\frac{L}{c}\frac{1}{\Delta S},
\eeq 
where
\beq 
\Delta p_{j,\perp}=\Delta \mathbfit{p}_{j}\cdot \mathbfit{n}
\eeq
is the projection of the transferred momentum on the local normal to the magnetospheric surface
\beq 
\mathbfit{n} = 
\left(\begin{array}{c} 
{\rm sgn}\lambda\,\sin(\lambda-\chi)\cos\varphi \\ 
{\rm sgn}\lambda\,\sin(\lambda-\chi)\sin\varphi \\
-{\rm sgn}\lambda\,\cos(\lambda-\chi)
\end {array}\right). 
\eeq

\section{Numerical results}
\label{sec:NumRes}

In our simulations, we explore the luminosity range of $10^{38}\,\ergs\lesssim L\lesssim 10^{40}\,\ergs$. The lower limit corresponds to the luminosity level where one would expect an optically thick magnetospheric accretion flow \citep{2017MNRAS.467.1202M}. 
Through Monte Carlo simulations, we investigate how the mass accretion rate and the multiple reprocessings of X-ray photons influence acceleration along magnetic field lines (Section \ref{sec:num_res_acc}) and pressure across field lines (Section \ref{sec:num_res_pressure}) due to the radiative force. 
By simulating multiple reprocessings of photons inside the magnetosphere, we can estimate the average photon escape time from the envelope (Section \ref{sec:num_res_esc_time}).

\subsection{Radiative transfer through a layer}

Each photon interaction with the magnetospheric surface is associated with either a penetration through the layer or a reflection back into the NS magnetosphere. 
The fraction of photons that penetrate through the layer, $f_{\rm out}$, is dependent on the optical thickness $\tau_{\rm e}$ and the incident angle of a photons (see Fig.\,\ref{pic:sc_f_out}). 
The larger the thickness and the angle between the local normal and initial photon momentum, the smaller the fraction of photons that penetrate through the layer. 
For the case of photons that initially move along the local normal, the fraction can be well approximated by (\ref{eq:f_out_approx}) if $\tau_{\rm e}\gg 1$ (compare black dotted line and the upper red solid line in  Fig.\,\ref{pic:sc_f_out}).
It is expected that the most of momentum along the field lines is transferred to the accretion flow within the layer of optical thickness $\sim 1$.
After the first scattering, the photons ``forget" the initial direction of their motion and contribute to the photon energy flux across magnetic field lines.

\subsection{Radiative force and acceleration along the field lines}
\label{sec:num_res_acc}

\begin{figure}
\centering 
\includegraphics[width=8.3cm]{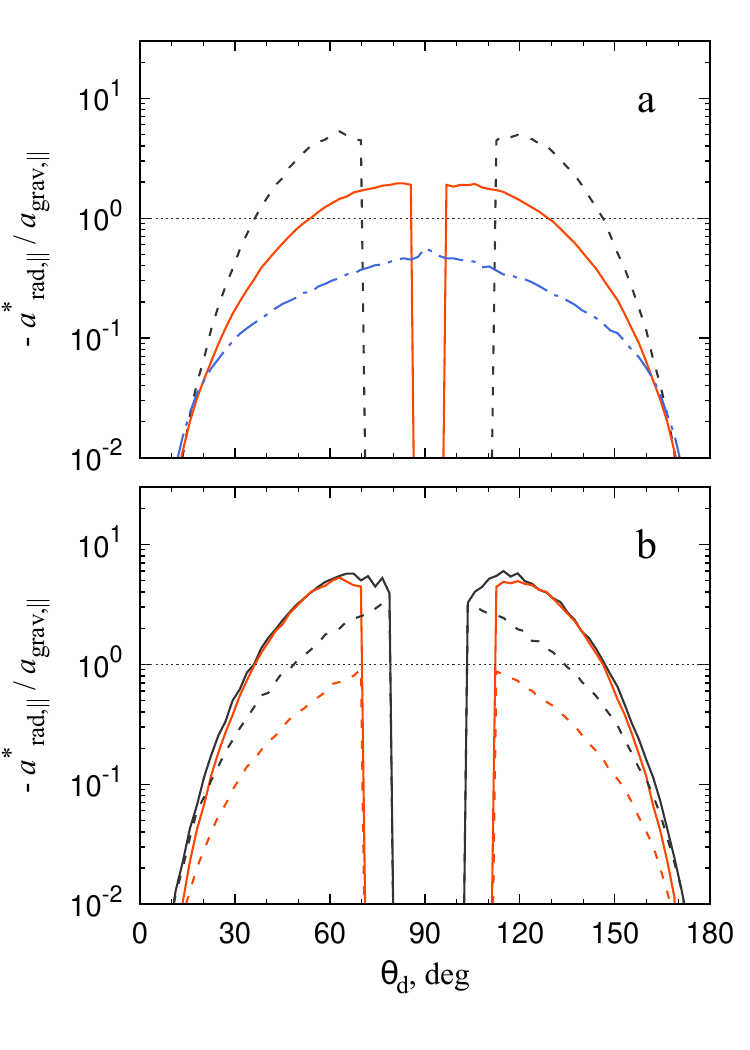}
\caption{
The ratio of the radiative to gravitational accelerations along the field lines as a function of coordinate $\theta_B$.
(a) 
Different lines correspond to different luminosities: 
$L=10^{38},\,10^{39},\, 10^{40}\,\ergs$ (blue dashed-dotted, solid red, and black dashed lines respectively).
(b)
The accretion luminosity is fixed at $10^{40}\,\ergs$.
Black (red) lines correspond to the surface field strength of $B=10^{14}\,{\rm G}$ ($B=3\times 10^{12}\,{\rm G}$).
The solid lines account for multiple reflection of X-ray photons while the dashed ones account for the first reflection only.
The dips at $\theta_B\sim 90^\circ$ appear due to the accretion disc that makes the effective surface density extremely large (see expression \ref{eq:a_par}). 
One can see that under the condition of sufficiently high luminosity, the acceleration along the field lines due to the radiative force can exceed the acceleration due to the gravitational force.
Note that the ratio is considered with a minus sign, and positive values indicate scenarios where the acceleration resulting from the radiative force opposes the acceleration due to gravity along the field lines.
}
\label{pic:sc_a_rad_2}
\end{figure}

The radiative force acting along magnetic field lines is proportional to the momentum transferred along the field lines (see \ref{eq:a_par} and \ref{eq:momentum_par}).
According to our simulations, the radiative acceleration along magnetic field lines is always in the opposite direction of the projection of the gravitational acceleration along magnetic field lines. 
A larger mass accretion rate and luminosity corresponds with a larger radiative force and acceleration along the field lines (see Fig.\,\ref{pic:sc_a_rad_2}a).
This occurs because there is a larger energy release at the NS surface, and a greater optical thickness of the accretion flow  results in a larger number of photon reflections (and hence, more momentum is transferred to the accreting material) before photons escape the system.
To estimate the contribution of the radiative force to the acceleration, we assume that the momentum of reprocessed photons is distributed homogeneously over the local optical thickness of the accretion flow (i.e. there is no gradient of radiative acceleration along $\tau_{\rm e}$).
Multiple reflections of X-ray photons affects the acceleration, especially in regions located close to the disc plane, and could increase the ratio $a^*_{\rm rad}/a_{\rm grav}$ by nearly an order of magnitude (see Fig.\,\ref{pic:sc_a_rad_2}b).
It is expected that in the case of $a^*_{\rm rad}/a_{\rm grav}>1$, the radiative force will begin to decelerate the accretion flow.
Remarkably, the distribution of the acceleration due to the radiative force becomes less sensitive to the specific beam pattern at higher mass accretion rates and luminosity.
It happens because at high mass accretion rates and larger optical thickness of accretion flow the radiative force is largely due to the photons that have already experienced one or a few reprocessing.

\subsection{Radiative pressure across the field lines}
\label{sec:num_res_pressure}

\begin{figure}
\centering 
\includegraphics[width=8.3cm]{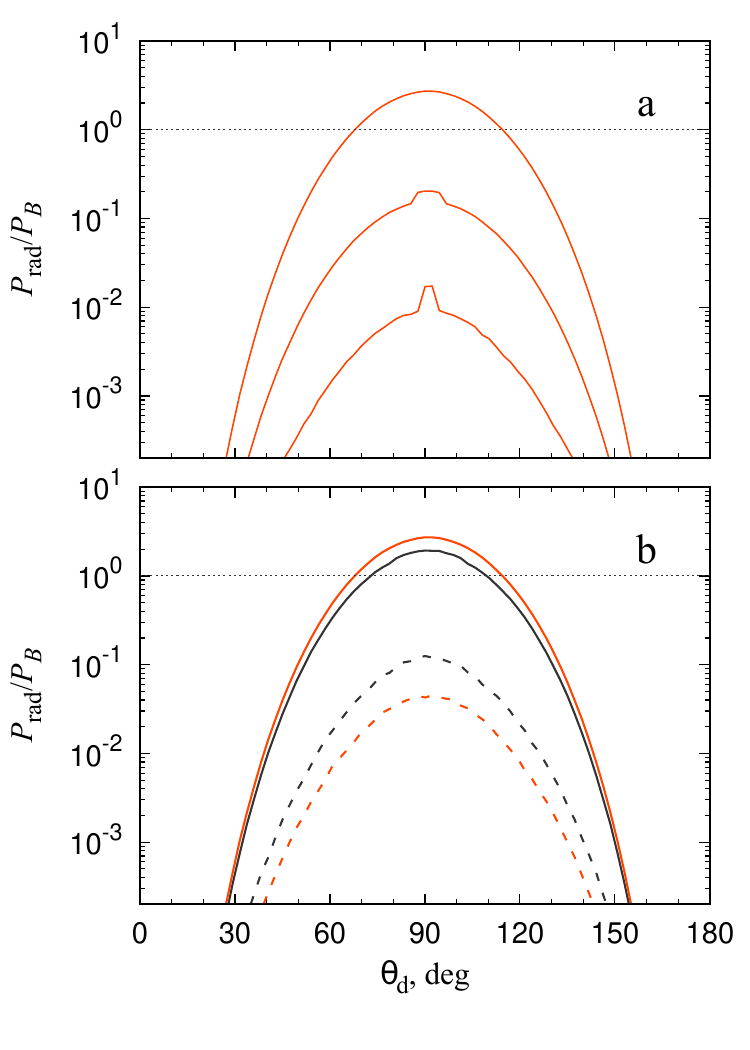}
\caption{
The ratio of the radiation pressure $P_{\rm rad}$ and the magnetic field pressure $P_B$ at the magnetospheric surface as a function of $\theta_B$ coordinate.
(a)
Different lines correspond to different accretion luminosities: $L=10^{38},\,10^{39},\,10^{40}\,\ergs$ (from bottom to top).
One can see that under condition of sufficiently high accretion luminosity, the radiative pressure can be larger than the local pressure of magnetic field (the field strength is fixed and is taken to be $B=3\times 10^{12}\,{\rm G}$ at the NS surface).
(b)
Solid lines show the result accounting for multiple reflections, while the dashed lines illustrate the case when only the first reflection is taken into account.
Red and black lines correspond to the case of surface magnetic field strength $3\times 10^{12}\,{\rm G}$ and $B=10^{14}\,{\rm G}$ respectively, luminosity is fixed at $L=10^{40}\,\ergs$.
}
\label{pic:sc_P_rad}
\end{figure}

Reflections of X-ray photons lead to momentum transfer across magnetic field lines, resulting in the anticipated radiative pressure. 
The radiative pressure across the field lines increases with luminosity and may surpass the local pressure of the magnetic field, denoted as $P_{B}=B^2/(8\pi)$, particularly in regions near the accretion disc plane, where the magnetic field is weaker (refer to Fig.\,\ref{pic:sc_P_rad}a).
When $P_{\rm rad}\gtrsim P_{\rm B}$, the geometry of the magnetic field lines can be influenced by the radiation pressure. 
Note that the local sharp maxima around $\theta_{\rm d}\approx 90\degr$ in the simulated curves, corresponding to $L=10^{38}\,\ergs$ and $10^{39}\,\ergs$ in Fig.\,\ref{pic:sc_P_rad}a, arise from the presence of the accretion disc, which reprocesses all photons reaching it.

The radiative pressure is significantly impacted by multiple reflections of X-ray photons, causing the pressure to increase by more than an order of magnitude (see Fig.\,\ref{pic:sc_P_rad}b).
Similarly to the radiate acceleration, the pressure becomes less sensitive to the specific beam pattern of higher mass accretion rates.

\subsection{Photon escape time}
\label{sec:num_res_esc_time}

\begin{figure}
\centering 
\includegraphics[width=8.5cm]{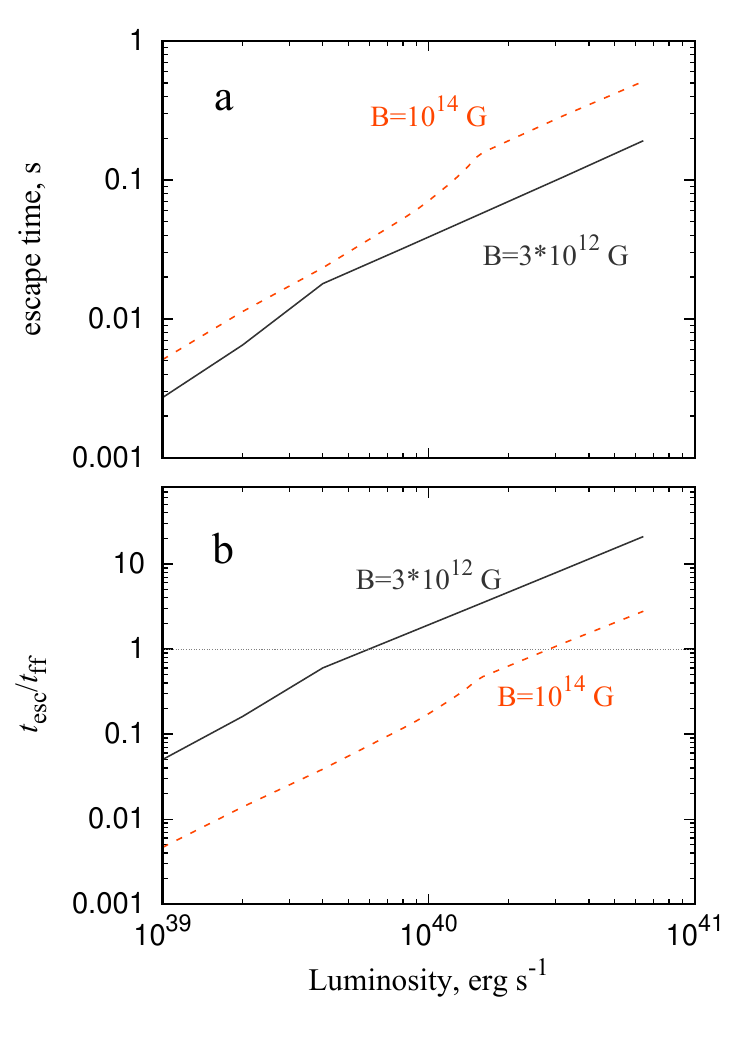}
\caption{
The typical escape time of X-ray photons from the magnetosphere of a NS covered by optically thick envelope in units of seconds (a), and in units of free-fall time from the inner radius of accretion disc (b).
Black and red lines illustrate the results for a surface magnetic field strength of $B=3\times 10^{12}\,{\rm G}$ and $10^{14}\,{\rm G}$, respectively.
The breaks in both curves are due to accretion disc transition from the geometrically thin to the geometrically thick state at the magnetospheric radius. 
}
\label{pic:sc_env_time}
\end{figure}

Photon escape time is affected by the mass accretion rate, as it determines the optical thickness of the flow and the geometrical size of the NS magnetosphere.  
The larger the mass accretion rate and luminosity, the longer the average time photons will travel within the magnetosphere (see Fig.\,\ref{pic:sc_env_time}).
The dependence of the average escape time on the mass accretion rate is approximately linear. 
Note, however, that according to \citealt{2019A&A...626A..18C}, the inner disc radius stabilizes and is weakly dependent on the mass accretion rate when it is advective or dominated by the radiation pressure (see Fig.\,26 in \citealt{2022arXiv220414185M}).
If it is indeed the case, the dependence of the photon travel time on the mass accretion rate should be stronger. 
The absolute value of the escape time tends to be longer in the case of a stronger magnetic field, due to a larger magnetosphere of a NS (see Eq.\,\ref{eq:Rm}).
The dependence of the escape time on the NS field strength is not strong because 
(i) the magnetospheric radius depend weakly on the field trength: $R_{\rm m}\propto B^{4/7}$, and 
(ii) the larger the magnetosphere, the smaller the optical thickness of the magnetospheric flow.
At an accretion luminosity of $\sim 10^{40}\,\ergs$, the photon escape time is $\sim 0.1\,{\rm s}$ (see Fig.\,\ref{pic:sc_env_time}a). 
 
The characteristic time of photon escape from the magnetosphere determines the time scale on which the radiation force and pressure are established. 
If the typical timescale of the accretion flow variations are smaller than (or about) the characteristic time of photon escape, the radiation field inside the magnetosphere is determined not only by the current configuration of the accretion flow, but also by the history of its variations.
At sufficiently high mass accretion rates, the escape time of X-ray photons can be longer than the typical time of free-fall from the inner radius of accretion disc (see Fig.\,\ref{pic:sc_env_time}b):
\beq
t_{\rm ff}\simeq 0.2\,m^{-1/2}\left(\frac{R_{\rm m}}{10^8\,{\rm cm}}\right)^{3/2}\,{\rm s},
\eeq
where $R_{\rm m}$ is given by (\ref{eq:Rm}).

It is expected that the escape time of photons from the magnetosphere affect features of aperiodic variability in XRPs.
The variability on timescales $t<t_{\rm esc}$ is expected to be suppressed (see, e.g., \citealt{2019MNRAS.484..687M}).

\section{Summary and discussion}
\label{sec:Summary}

We have analyzed interactions of photons with the magnetospheric accretion flow in bright XRPs ($L\gtrsim 10^{38}\,\ergs$) including ULXs powered by accretion onto a NS.
In these objects, the accretion flow covering NS magnetosphere is expected to be optically thick due to the photons scattering by electrons and influence spectra and pulse profile formation \citep{2017MNRAS.467.1202M,2023MNRAS.525.4176B}.
Using precalculated maps of material distribution over the magnetospheric surface \citep{2019MNRAS.484..687M} and applying Monte Carlo simulations, we have investigated the distribution of radiative force along magnetic field lines and radiation pressure across the magnetic field lines over the magnetospheric surface.
We have estimated the typical time scale of photon escape from the magnetosphere of a NS at different mass accretion rates.
Our conclusions are: 
\begin{itemize}[leftmargin=*]
\item
Multiple reflections of X-ray photons increase (sometimes by more than an order of magnitude) both the radiative force along magnetic field lines (see Fig.\,\ref{pic:sc_a_rad_2}b) and the radiation pressure across the field lines (see Fig.\,\ref{pic:sc_P_rad}b).
\item
At sufficiently high luminosities, the magnitude of acceleration along the magnetic field lines due to the radiative force can exceed the acceleration along the field lines due to the gravitational force, particularly in regions near the equatorial plane of the magnetic dipole (see Fig.\,\ref{pic:sc_a_rad_2}a).
In these regions one would expect significant deceleration of accretion flow by the radiative force.
\item 
At an accretion luminosity of $L\gtrsim 5\times 10^{39}\,\ergs$, the radiative pressure across magnetic field lines can exceed the local magnetic field pressure (see Fig.\,\ref{pic:sc_P_rad}a).
The ratio $P_{\rm rad}/P_B$ tends to be larger in the regions located close to the equator of the magnetic dipole. 
In the case of $P_{\rm rad}/P_B>1$ one would expect local disruption of the accretion flow and possibly radiation driven mass losses.
\item 
We have demonstrated that increasing the mass accretion rate and luminosity results in a larger average escape time for photons inside the magnetosphere (see Fig.\,\ref{pic:sc_env_time}). 
The average time of photon escape depends on the mass accretion rate and size of NS magnetosphere (i.e., on the NS magnetic field strength and structure).
The dependence of the photon escape time on the mass accretion rate is approximately linear, while the dependence on the stellar field strength is weak.
At accretion luminosity $L\approx 10^{40}\,\ergs$, the avarage photon escape time $\sim 0.1\,{\rm s}$.
The typical time scale of photon escape determines the time scale of radiative force and pressure establishing, which has to be taken into account in calculations of magnetosphere accretion flow dynamics.
\end{itemize}

The significant radiative force along magnetic field lines, not previously considered in earlier publications \citep{2017MNRAS.467.1202M,2019MNRAS.484..687M,2023MNRAS.525.4176B}, can profoundly impact the dynamics of the accretion flow within the magnetospheric radius. In this scenario, the analysis of material distribution over the magnetospheric surface, its influence on spectra formation, and even the stability of the accretion flow must account for the effects of the radiative force.
As the radiative acceleration along magnetic field lines can surpass the acceleration due to gravitational force, the accretion flow towards the neutron star surface may be impeded by radiation. 
This, in turn, results in a decrease in accretion luminosity, a reduction in radiative force, and the subsequent resumption of the accretion process. Consequently, the appearance of oscillations in mass accretion rate within the magnetosphere is anticipated \citep{2024arXiv240212965M}.

In line with previous findings \citep{2017MNRAS.467.1202M}, the radiation pressure on the accretion flow can surpass the local magnetic field pressure. 
The radiation pressure has the potential to distort magnetic field lines, thereby influencing the geometry and dynamics of the magnetospheric accretion flow.

The extended timescale for photon escape from the envelope at high mass accretion rates and luminosities (see Fig.\,\ref{pic:sc_env_time}) is anticipated to influence the timing properties of X-ray aperiodic variability. 
Strong suppression is expected in the power density spectra on timescales shorter than the average time of photon escape (see \citealt{2019MNRAS.484..687M} for detailed discussions).
Given that the typical free-fall time from the inner disc radius in X-ray pulsars is approximately $\sim 0.1\,{\rm s}$ (comparable to the timescale of photon escape during high luminosity states) and the accretion flow can be rendered unstable due to the influence of radiative force and pressure \citep{2024arXiv240212965M}, the structure of the magnetospheric flow and radiative field becomes coupled. 
Under these conditions, their temporal evolution cannot be calculated separately in numerical models.

\section*{Acknowledgements}

AAM thanks UKRI Stephen Hawking fellowship.
We are grateful to Simon Portegies Zwart for discussions and
our anonymous referee for useful comments and suggestions which helped us improve the paper.

\section*{Data availability}

The calculations presented in this paper were performed using a private code developed and owned by the corresponding author. All the data appearing in the figures are available upon request. 


\begin{thebibliography}{}
\makeatletter
\relax
\def\mn@urlcharsother{\let\do\@makeother \do\$\do\&\do\#\do\^\do\_\do\%\do\~}
\def\mn@doi{\begingroup\mn@urlcharsother \@ifnextchar [ {\mn@doi@}
  {\mn@doi@[]}}
\def\mn@doi@[#1]#2{\def\@tempa{#1}\ifx\@tempa\@empty \href
  {http://dx.doi.org/#2} {doi:#2}\else \href {http://dx.doi.org/#2} {#1}\fi
  \endgroup}
\def\mn@eprint#1#2{\mn@eprint@#1:#2::\@nil}
\def\mn@eprint@arXiv#1{\href {http://arxiv.org/abs/#1} {{\tt arXiv:#1}}}
\def\mn@eprint@dblp#1{\href {http://dblp.uni-trier.de/rec/bibtex/#1.xml}
  {dblp:#1}}
\def\mn@eprint@#1:#2:#3:#4\@nil{\def\@tempa {#1}\def\@tempb {#2}\def\@tempc
  {#3}\ifx \@tempc \@empty \let \@tempc \@tempb \let \@tempb \@tempa \fi \ifx
  \@tempb \@empty \def\@tempb {arXiv}\fi \@ifundefined
  {mn@eprint@\@tempb}{\@tempb:\@tempc}{\expandafter \expandafter \csname
  mn@eprint@\@tempb\endcsname \expandafter{\@tempc}}}

\bibitem[\protect\citeauthoryear{{Arons} \& {Lea}}{{Arons} \&
  {Lea}}{1976}]{1976ApJ...207..914A}
{Arons} J.,  {Lea} S.~M.,  1976, \mn@doi [\apj] {10.1086/154562}, \href
  {https://ui.adsabs.harvard.edu/abs/1976ApJ...207..914A} {207, 914}

\bibitem[\protect\citeauthoryear{{Bachetti} et~al.,}{{Bachetti}
  et~al.}{2014}]{2014Natur.514..202B}
{Bachetti} M.,  et~al., 2014, \mn@doi [\nat] {10.1038/nature13791}, \href
  {https://ui.adsabs.harvard.edu/abs/2014Natur.514..202B} {514, 202}

\bibitem[\protect\citeauthoryear{{Basko} \& {Sunyaev}}{{Basko} \&
  {Sunyaev}}{1976}]{1976MNRAS.175..395B}
{Basko} M.~M.,  {Sunyaev} R.~A.,  1976, \mn@doi [\mnras]
  {10.1093/mnras/175.2.395}, \href
  {https://ui.adsabs.harvard.edu/abs/1976MNRAS.175..395B} {175, 395}

\bibitem[\protect\citeauthoryear{{Brice}, {Zane}, {Taverna}, {Turolla}  \&
  {Wu}}{{Brice} et~al.}{2023}]{2023MNRAS.525.4176B}
{Brice} N.,  {Zane} S.,  {Taverna} R.,  {Turolla} R.,   {Wu} K.,  2023, \mn@doi
  [\mnras] {10.1093/mnras/stad2391}, \href
  {https://ui.adsabs.harvard.edu/abs/2023MNRAS.525.4176B} {525, 4176}

\bibitem[\protect\citeauthoryear{{Burnard}, {Arons}  \& {Lea}}{{Burnard}
  et~al.}{1983}]{1983ApJ...266..175B}
{Burnard} D.~J.,  {Arons} J.,   {Lea} S.~M.,  1983, \mn@doi [\apj]
  {10.1086/160768}, \href
  {https://ui.adsabs.harvard.edu/abs/1983ApJ...266..175B} {266, 175}

\bibitem[\protect\citeauthoryear{{Carpano}, {Haberl}, {Maitra}  \&
  {Vasilopoulos}}{{Carpano} et~al.}{2018}]{2018MNRAS.476L..45C}
{Carpano} S.,  {Haberl} F.,  {Maitra} C.,   {Vasilopoulos} G.,  2018, \mn@doi
  [\mnras] {10.1093/mnrasl/sly030}, \href
  {https://ui.adsabs.harvard.edu/abs/2018MNRAS.476L..45C} {476, L45}

\bibitem[\protect\citeauthoryear{{Chashkina}, {Lipunova}, {Abolmasov}  \&
  {Poutanen}}{{Chashkina} et~al.}{2019}]{2019A&A...626A..18C}
{Chashkina} A.,  {Lipunova} G.,  {Abolmasov} P.,   {Poutanen} J.,  2019,
  \mn@doi [\aap] {10.1051/0004-6361/201834414}, \href
  {https://ui.adsabs.harvard.edu/abs/2019A&A...626A..18C} {626, A18}

\bibitem[\protect\citeauthoryear{{Fabrika}, {Atapin}, {Vinokurov}  \&
  {Sholukhova}}{{Fabrika} et~al.}{2021}]{2021AstBu..76....6F}
{Fabrika} S.~N.,  {Atapin} K.~E.,  {Vinokurov} A.~S.,   {Sholukhova} O.~N.,
  2021, \mn@doi [Astrophysical Bulletin] {10.1134/S1990341321010077}, \href
  {https://ui.adsabs.harvard.edu/abs/2021AstBu..76....6F} {76, 6}

\bibitem[\protect\citeauthoryear{{Frank}, {King}  \& {Raine}}{{Frank}
  et~al.}{2002}]{2002apa..book.....F}
{Frank} J.,  {King} A.,   {Raine} D.~J.,  2002, {Accretion Power in
  Astrophysics: Third Edition}

\bibitem[\protect\citeauthoryear{{F{\"u}rst} et~al.,}{{F{\"u}rst}
  et~al.}{2016}]{2016ApJ...831L..14F}
{F{\"u}rst} F.,  et~al., 2016, \mn@doi [\apjl] {10.3847/2041-8205/831/2/L14},
  \href {https://ui.adsabs.harvard.edu/abs/2016ApJ...831L..14F} {831, L14}

\bibitem[\protect\citeauthoryear{{Ghosh} \& {Lamb}}{{Ghosh} \&
  {Lamb}}{1978}]{1978ApJ...223L..83G}
{Ghosh} P.,  {Lamb} F.~K.,  1978, \mn@doi [\apjl] {10.1086/182734}, \href
  {https://ui.adsabs.harvard.edu/abs/1978ApJ...223L..83G} {223, L83}

\bibitem[\protect\citeauthoryear{{Ghosh} \& {Lamb}}{{Ghosh} \&
  {Lamb}}{1979}]{1979ApJ...234..296G}
{Ghosh} P.,  {Lamb} F.~K.,  1979, \mn@doi [\apj] {10.1086/157498}, \href
  {https://ui.adsabs.harvard.edu/abs/1979ApJ...234..296G} {234, 296}

\bibitem[\protect\citeauthoryear{{Illarionov} \& {Sunyaev}}{{Illarionov} \&
  {Sunyaev}}{1975}]{1975A&A....39..185I}
{Illarionov} A.~F.,  {Sunyaev} R.~A.,  1975, \aap, \href
  {https://ui.adsabs.harvard.edu/abs/1975A&A....39..185I} {39, 185}

\bibitem[\protect\citeauthoryear{{Israel} et~al.,}{{Israel}
  et~al.}{2017a}]{2017Sci...355..817I}
{Israel} G.~L.,  et~al., 2017a, \mn@doi [Science] {10.1126/science.aai8635},
  \href {https://ui.adsabs.harvard.edu/abs/2017Sci...355..817I} {355, 817}

\bibitem[\protect\citeauthoryear{{Israel} et~al.,}{{Israel}
  et~al.}{2017b}]{2017MNRAS.466L..48I}
{Israel} G.~L.,  et~al., 2017b, \mn@doi [\mnras] {10.1093/mnrasl/slw218}, \href
  {https://ui.adsabs.harvard.edu/abs/2017MNRAS.466L..48I} {466, L48}

\bibitem[\protect\citeauthoryear{{King}, {Lasota}  \& {Middleton}}{{King}
  et~al.}{2023}]{2023NewAR..9601672K}
{King} A.,  {Lasota} J.-P.,   {Middleton} M.,  2023, \mn@doi [\nar]
  {10.1016/j.newar.2022.101672}, \href
  {https://ui.adsabs.harvard.edu/abs/2023NewAR..9601672K} {96, 101672}

\bibitem[\protect\citeauthoryear{{Koliopanos}, {Vasilopoulos}, {Buchner},
  {Maitra}  \& {Haberl}}{{Koliopanos} et~al.}{2019}]{2019A&A...621A.118K}
{Koliopanos} F.,  {Vasilopoulos} G.,  {Buchner} J.,  {Maitra} C.,   {Haberl}
  F.,  2019, \mn@doi [\aap] {10.1051/0004-6361/201834144}, \href
  {https://ui.adsabs.harvard.edu/abs/2019A&A...621A.118K} {621, A118}

\bibitem[\protect\citeauthoryear{{Kong} et~al.,}{{Kong}
  et~al.}{2022}]{2022ApJ...933L...3K}
{Kong} L.-D.,  et~al., 2022, \mn@doi [\apjl] {10.3847/2041-8213/ac7711}, \href
  {https://ui.adsabs.harvard.edu/abs/2022ApJ...933L...3K} {933, L3}

\bibitem[\protect\citeauthoryear{{Kulkarni} \& {Romanova}}{{Kulkarni} \&
  {Romanova}}{2008}]{2008MNRAS.386..673K}
{Kulkarni} A.~K.,  {Romanova} M.~M.,  2008, \mn@doi [\mnras]
  {10.1111/j.1365-2966.2008.13094.x}, \href
  {https://ui.adsabs.harvard.edu/abs/2008MNRAS.386..673K} {386, 673}

\bibitem[\protect\citeauthoryear{{M{\"o}nkk{\"o}nen}, {Tsygankov}, {Mushtukov},
  {Doroshenko}, {Suleimanov}  \& {Poutanen}}{{M{\"o}nkk{\"o}nen}
  et~al.}{2022}]{2022MNRAS.515..571M}
{M{\"o}nkk{\"o}nen} J.,  {Tsygankov} S.~S.,  {Mushtukov} A.~A.,  {Doroshenko}
  V.,  {Suleimanov} V.~F.,   {Poutanen} J.,  2022, \mn@doi [\mnras]
  {10.1093/mnras/stac1828}, \href
  {https://ui.adsabs.harvard.edu/abs/2022MNRAS.515..571M} {515, 571}

\bibitem[\protect\citeauthoryear{{Mushtukov} \& {Tsygankov}}{{Mushtukov} \&
  {Tsygankov}}{2022}]{2022arXiv220414185M}
{Mushtukov} A.,  {Tsygankov} S.,  2022, \mn@doi [arXiv e-prints]
  {10.48550/arXiv.2204.14185}, \href
  {https://ui.adsabs.harvard.edu/abs/2022arXiv220414185M} {p. arXiv:2204.14185}

\bibitem[\protect\citeauthoryear{{Mushtukov}, {Suleimanov}, {Tsygankov}  \&
  {Poutanen}}{{Mushtukov} et~al.}{2015}]{2015MNRAS.447.1847M}
{Mushtukov} A.~A.,  {Suleimanov} V.~F.,  {Tsygankov} S.~S.,   {Poutanen} J.,
  2015, \mn@doi [\mnras] {10.1093/mnras/stu2484}, \href
  {https://ui.adsabs.harvard.edu/abs/2015MNRAS.447.1847M} {447, 1847}

\bibitem[\protect\citeauthoryear{{Mushtukov}, {Suleimanov}, {Tsygankov}  \&
  {Ingram}}{{Mushtukov} et~al.}{2017}]{2017MNRAS.467.1202M}
{Mushtukov} A.~A.,  {Suleimanov} V.~F.,  {Tsygankov} S.~S.,   {Ingram} A.,
  2017, \mn@doi [\mnras] {10.1093/mnras/stx141}, \href
  {https://ui.adsabs.harvard.edu/abs/2017MNRAS.467.1202M} {467, 1202}

\bibitem[\protect\citeauthoryear{{Mushtukov}, {Ingram}, {Middleton}, {Nagirner}
   \& {van der Klis}}{{Mushtukov} et~al.}{2019}]{2019MNRAS.484..687M}
{Mushtukov} A.~A.,  {Ingram} A.,  {Middleton} M.,  {Nagirner} D.~I.,   {van der
  Klis} M.,  2019, \mn@doi [\mnras] {10.1093/mnras/sty3525}, \href
  {https://ui.adsabs.harvard.edu/abs/2019MNRAS.484..687M} {484, 687}

\bibitem[\protect\citeauthoryear{{Mushtukov}, {Ingram}, {Suleimanov},
  {DiLullo}, {Middleton}, {Tsygankov}, {van der Klis}  \& {Portegies
  Zwart}}{{Mushtukov} et~al.}{2024}]{2024arXiv240212965M}
{Mushtukov} A.~A.,  {Ingram} A.,  {Suleimanov} V.~F.,  {DiLullo} N.,
  {Middleton} M.,  {Tsygankov} S.~S.,  {van der Klis} M.,   {Portegies Zwart}
  S.,  2024, \mn@doi [arXiv e-prints] {10.48550/arXiv.2402.12965}, \href
  {https://ui.adsabs.harvard.edu/abs/2024arXiv240212965M} {p. arXiv:2402.12965}

\bibitem[\protect\citeauthoryear{{Nagirner}, {Milanova}, {Dementyev}  \&
  {Volkov}}{{Nagirner} et~al.}{2022}]{2022Ap.....65..560N}
{Nagirner} D.~I.,  {Milanova} Y.~V.,  {Dementyev} A.~V.,   {Volkov} E.~V.,
  2022, \mn@doi [Astrophysics] {10.1007/s10511-023-09764-4}, \href
  {https://ui.adsabs.harvard.edu/abs/2022Ap.....65..560N} {65, 560}

\bibitem[\protect\citeauthoryear{{Psaltis} \& {Chakrabarty}}{{Psaltis} \&
  {Chakrabarty}}{1999}]{1999ApJ...521..332P}
{Psaltis} D.,  {Chakrabarty} D.,  1999, \mn@doi [\apj] {10.1086/307525}, \href
  {https://ui.adsabs.harvard.edu/abs/1999ApJ...521..332P} {521, 332}

\bibitem[\protect\citeauthoryear{{Reig}}{{Reig}}{2011}]{2011Ap&SS.332....1R}
{Reig} P.,  2011, \mn@doi [\apss] {10.1007/s10509-010-0575-8}, \href
  {https://ui.adsabs.harvard.edu/abs/2011Ap&SS.332....1R} {332, 1}

\bibitem[\protect\citeauthoryear{{Rodr{\'\i}guez Castillo}
  et~al.,}{{Rodr{\'\i}guez Castillo} et~al.}{2020}]{2020ApJ...895...60R}
{Rodr{\'\i}guez Castillo} G.~A.,  et~al., 2020, \mn@doi [\apj]
  {10.3847/1538-4357/ab8a44}, \href
  {https://ui.adsabs.harvard.edu/abs/2020ApJ...895...60R} {895, 60}

\bibitem[\protect\citeauthoryear{{Romanova}, {Ustyugova}, {Koldoba}, {Wick}  \&
  {Lovelace}}{{Romanova} et~al.}{2003}]{2003ApJ...595.1009R}
{Romanova} M.~M.,  {Ustyugova} G.~V.,  {Koldoba} A.~V.,  {Wick} J.~V.,
  {Lovelace} R.~V.~E.,  2003, \mn@doi [\apj] {10.1086/377514}, \href
  {https://ui.adsabs.harvard.edu/abs/2003ApJ...595.1009R} {595, 1009}

\bibitem[\protect\citeauthoryear{{Romanova}, {Ustyugova}, {Koldoba}  \&
  {Lovelace}}{{Romanova} et~al.}{2004}]{2004ApJ...610..920R}
{Romanova} M.~M.,  {Ustyugova} G.~V.,  {Koldoba} A.~V.,   {Lovelace} R.~V.~E.,
  2004, \mn@doi [\apj] {10.1086/421867}, \href
  {https://ui.adsabs.harvard.edu/abs/2004ApJ...610..920R} {610, 920}

\bibitem[\protect\citeauthoryear{{Sathyaprakash} et~al.,}{{Sathyaprakash}
  et~al.}{2019}]{2019MNRAS.488L..35S}
{Sathyaprakash} R.,  et~al., 2019, \mn@doi [\mnras] {10.1093/mnrasl/slz086},
  \href {https://ui.adsabs.harvard.edu/abs/2019MNRAS.488L..35S} {488, L35}

\bibitem[\protect\citeauthoryear{{Shakura} \& {Sunyaev}}{{Shakura} \&
  {Sunyaev}}{1973}]{1973A&A....24..337S}
{Shakura} N.~I.,  {Sunyaev} R.~A.,  1973, \aap, \href
  {https://ui.adsabs.harvard.edu/abs/1973A&A....24..337S} {24, 337}

\bibitem[\protect\citeauthoryear{{Shakura}, {Postnov}  \&
  {Prokhorov}}{{Shakura} et~al.}{1991}]{1991PAZh...17..803S}
{Shakura} N.~I.,  {Postnov} K.~A.,   {Prokhorov} M.~E.,  1991, Pisma v
  Astronomicheskii Zhurnal, \href
  {https://ui.adsabs.harvard.edu/abs/1991PAZh...17..803S} {17, 803}

\bibitem[\protect\citeauthoryear{{Staubert} et~al.,}{{Staubert}
  et~al.}{2019}]{2019A&A...622A..61S}
{Staubert} R.,  et~al., 2019, \mn@doi [\aap] {10.1051/0004-6361/201834479},
  \href {https://ui.adsabs.harvard.edu/abs/2019A&A...622A..61S} {622, A61}

\bibitem[\protect\citeauthoryear{{Suleimanov}, {Lipunova}  \&
  {Shakura}}{{Suleimanov} et~al.}{2007}]{2007ARep...51..549S}
{Suleimanov} V.~F.,  {Lipunova} G.~V.,   {Shakura} N.~I.,  2007, \mn@doi
  [Astronomy Reports] {10.1134/S1063772907070049}, \href
  {https://ui.adsabs.harvard.edu/abs/2007ARep...51..549S} {51, 549}

\bibitem[\protect\citeauthoryear{{Sunyaev}}{{Sunyaev}}{1976}]{1976SvAL....2..111S}
{Sunyaev} R.~A.,  1976, Soviet Astronomy Letters, \href
  {https://ui.adsabs.harvard.edu/abs/1976SvAL....2..111S} {2, 111}

\bibitem[\protect\citeauthoryear{{Tsygankov}, {Lutovinov}, {Doroshenko},
  {Mushtukov}, {Suleimanov}  \& {Poutanen}}{{Tsygankov}
  et~al.}{2016}]{2016A&A...593A..16T}
{Tsygankov} S.~S.,  {Lutovinov} A.~A.,  {Doroshenko} V.,  {Mushtukov} A.~A.,
  {Suleimanov} V.,   {Poutanen} J.,  2016, \mn@doi [\aap]
  {10.1051/0004-6361/201628236}, \href
  {https://ui.adsabs.harvard.edu/abs/2016A&A...593A..16T} {593, A16}

\bibitem[\protect\citeauthoryear{{Tsygankov}, {Doroshenko}, {Lutovinov},
  {Mushtukov}  \& {Poutanen}}{{Tsygankov} et~al.}{2017}]{2017A&A...605A..39T}
{Tsygankov} S.~S.,  {Doroshenko} V.,  {Lutovinov} A.~A.,  {Mushtukov} A.~A.,
  {Poutanen} J.,  2017, \mn@doi [\aap] {10.1051/0004-6361/201730553}, \href
  {https://ui.adsabs.harvard.edu/abs/2017A&A...605A..39T} {605, A39}

\makeatother
\end{thebibliography}

\appendix

\section{ON THE FRACTION OF PHOTONS PENETRATING THROUGH THE LAYER}
\label{app:fraction}

In the case of a layer illuminated by photons, the photons can either penetrate through, be absorbed, or be reflected by the layer. 
The probabilities of these different scenarios depend on the optical thickness of the layer, as well as the nature of opacity resulting from the scattering and absorption processes.

Consider a plane-parallel layer with a given optical thickness, denoted as $\tau_e$. Assume that isotropic scattering, without photon absorption (i.e., conservative scattering), is the sole process that governs photon interaction within the layer. Under this condition, photons are either reflected or penetrate through the layer, and the total number of photons is conserved.
The fraction of photons that penetrate through the layer depends on both $\tau_e$ and the initial direction of the photon momentum relative to the normal to the layer, represented by the angle $\theta_{\rm in}$. For the specific case where $\theta_{\rm in}=0$, there exists an approximate expression (\ref{eq:f_out_approx}) for the fraction of photons capable of penetrating through the layer. For arbitrary angles $\theta_{\rm in}$, the fraction of  photons that penetrate through the layer can be determined numerically.
To calculate the fraction of photons that penetrate through the layer, we conduct Monte Carlo simulations using the following algorithm, which can be divided into several steps:
\begin{enumerate}[leftmargin=*]
\item\label{ph_ini} Initiate a photon with a given initial direction of motion, determined by the angle between the momentum and the normal, denoted as $\theta$, from a boundary of a layer (i.e., at optical depth $\tau_{\rm ini}=0$).
\item\label{free_path} Generate a random number $X_1$ to determine the free path of a photon inside a layer in units of mean free path:
\beq 
\Delta\tau = -\ln X_1.
\eeq 
\item Use the obtained free path of the photon to determine the optical depth of the next scattering event:
\beq 
\tau_{\rm fin}=\tau_{\rm ini}+\Delta\tau\,\cos\theta.
\eeq
If $\tau_{\rm fin}<0$ ($>\tau_e$), the photon is reflected (penetrates through) the layer. 
In this case, we account for the photon and return to step \ref{ph_ini}.
If $0<\tau_{\rm fin}<\tau_e$, the photon undergoes the next scattering inside the layer, and proceed to step \ref{redirection}.
\item\label{redirection} 
Set the optical depth of reemitted photons as $\tau_{\rm ini}=\tau_{\rm fin}$.
Re-emit the photon isotropically, and calculate a new direction of photon motion:
\beq 
\theta = {\rm arccos}(1-2X_2),
\eeq 
where $X_2$ is a random number. 
Return to step \ref{free_path}.
\end{enumerate}

The fraction of photons that penetrate through a layer is 
\beq 
f_{\rm out}\approx\frac{N_{\rm out}}{N_{\rm tot}},
\eeq 
where $N_{\rm tot}$ is a total number of photons in a simulation, and $N_{\rm out}$ is a number of photons penetrated through a layer.
Results of numerical simulations are shown in Fig.\,\ref{pic:sc_f_out}.
The larger the optical thickness and angle between the normal and photon momentum, the smaller the fraction of photons penetrating through the layer. 
At $\tau_e\gg 1$, $f_{\rm out}\propto \tau_e^{-1}$.
For $\tau_e>50$ and $\theta<1.5\,{\rm rad}$, one can use approximation
\beq 
f_{\rm out}(\tau_e,\theta)\approx
f_{\rm out}(\tau_e,\theta=0)-\frac{\theta^2}{2.2\tau_e}.
\eeq 

\bsp 
\label{lastpage}
\end{document}